\pdfoutput=1
\documentclass[aip,reprint]{revtex4-1}

\usepackage{graphicx}
\usepackage{amsmath}
\usepackage[defaultmathsizes]{mathastext}

\begin{document}

\title{Radio frequency reflectometry and charge sensing of a precision placed donor in silicon}

\author{Samuel J Hile}
 \email{samhile@gmail.com}
\author{Matthew G House}%
\author{Eldad Peretz}%
\author{Jan Verduijn}%
\author{Daniel Widmann}%
\author{Takashi Kobayashi}%
\author{Sven Rogge}%
\author{Michelle Y Simmons}%
 \email{michelle.simmons@unsw.edu.au}
\affiliation{%
Centre for Quantum Computation and Communication Technology ($CQC^2T$),\\ School of Physics, University of New South Wales, Sydney 2052, Australia
}%



\begin{abstract}
We compare charge transitions on a deterministic single P donor in silicon using radio frequency reflectometry measurements with a tunnel coupled reservoir and DC charge sensing using a capacitively coupled single electron transistor (SET). 
By measuring the conductance through the SET and comparing this with the phase shift of the reflected RF excitation from the reservoir, we can discriminate between charge transfer within the SET channel and tunneling between the donor and reservoir. The RF measurement allows observation of donor electron transitions at every charge degeneracy point in contrast to the SET conductance signal where charge transitions are only observed at triple points. 
The tunnel coupled reservoir has the advantage of a large effective lever arm (${\sim}35\%$) allowing us to independently extract a neutral donor charging energy ${\sim}62\pm 17meV$. 
These results demonstrate that we can replace three terminal transistors by a single terminal dispersive reservoir, promising for high bandwidth scalable donor control and readout.
\end{abstract}

\maketitle

Phosphorus donor nuclear spins in silicon \cite{kane,zwanenburg,plaN} provide an excellent  platform for quantum computation with coherence times $>30$ seconds, and single qubit gate fidelities above $99.99\%$ in isotopically purified silicon\cite{muhonen}. 
Interaction with the nuclear spin occurs via the hyperfine interaction with the donor electron spin bound by the donor Coulomb potential. This bound electron represents a spin qubit in its own right \cite{plaE, hill}, 
with coherence times $>0.5$ seconds and gate fidelities above $99\%$ in isotopically pure silicon \cite{muhonen}. 
Measurement of single electron spin states has to date largely relied on spin to charge conversion followed by charge state readout through either a charge-sensing single electron transistor (SET) \cite{morello, buch} or quantum point contact (QPC) \cite{elzerman}. 
Both the SET and QPC necessarily require source and drain contacts and typically an additional gate to tune them to a sensitive operating point for high fidelity spin readout. Compared to the donor qubit such three-terminal read-out infrastructure requires a significant amount of on-chip space. 
Thus despite a viable path to scalable qubit architectures \cite{hollenberg}, one of the major challenges in scaling atomic-scale qubits is integrating enough readout transistors into a large-scale array of donor qubits where the donor separation may be as small as 10-20nm\cite{hollenberg}.

An alternate way to perform spin read-out is to use a radio frequency SET (RF-SET). Here an AC voltage is reflected off a resonant circuit which measures the AC conductance of an SET channel, thus giving information on the charge state of interest \cite{schoelkopf}.
Whilst RF-SET  reflectometry still requires three terminals, it effectively filters low frequency noise, such as charge noise and inductively coupled current noise, producing charge sensitivities at $10^{-5} e / \sqrt{Hz}$ with megahertz bandwidth in silicon\cite{schoelkopf,angus}. 
A  recent evolution of this technique is the single terminal dispersive gate sensor which instead monitors the AC impedance seen by a single gate\cite{colless} or lead\cite{persson} in the presence of nearby electron motion, which is now approaching the sensitivity seen in RF-SETs\cite{gonzalez}.
Reflectometry techniques have recently been applied to devices with randomly implanted donors, providing information on the location and coupling strength of the donor within the nanostructure beyond that accessible with direct transport techniques\cite{verduijn, villis14}.

\begin{figure}[!t]
\includegraphics[width=0.45\textwidth]{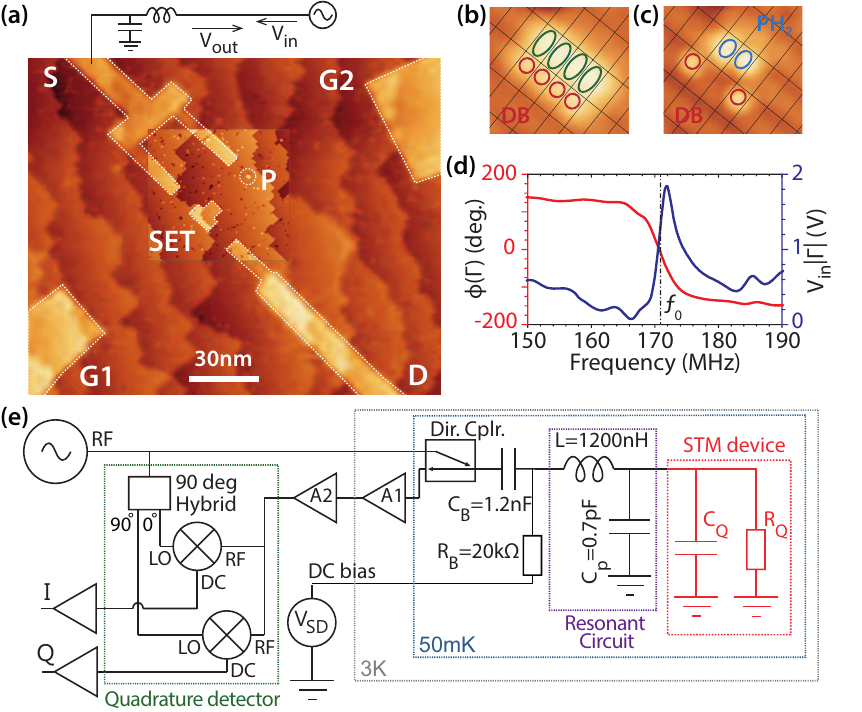}
\caption{\label{fig:one} \textbf{Device and circuit layout to compare charge sensing with an SET and reflectometry using a reservoir of a single P donor in silicon.}
(a) An STM image of the device, showing a hydrogen terminated silicon surface where hydrogen atoms have been removed with an STM tip to form the template for the creation of a single P donor capacitively coupled to an SET and tunnel coupled to the source reservoir. 
(b) STM image of the single donor incorporation site overlaid with the Si(2x1) surface atomic lattice grid, showing four desorbed adjacent dimers (green) before dosing and 
(c) after $PH_3$ dosing, showing two $PH_2$ fragments (blue). Red circles indicate single nonreactive dangling bonds.
(d) Amplitude and phase of the reflected signal around the LC resonance at $170.9MHz$ measured with $V_{SD} = V_{G1} = V_{G2} = 0$.  
(e) Schematic of the RF measurement circuit, showing the applied RF signal injected through a directional coupler, the STM device and resonant circuit at mK, followed by amplifiers at the 3K stage and at room temperature (A1, A2) and a quadrature detection circuit.} 
\end{figure}

In this paper we present RF characterisation of a precision placed single donor within a device where we can directly compare charge sensing of the donor using an SET with reflectometry using a single terminal that acts as a combined electron reservoir, control gate and readout sensor. 
Such a device allows us to confirm the presence of the single donor with conventional charge sensing and definitively distinguish this from other charge transitions within the device. 
A significant advantage of the dispersive measurement is that it only requires a single gate reservoir, allowing us to demonstrate the viability of this technique to form a small-footprint, scalable readout device for single donor electronics.

Figure 1(a) shows a scanning tunneling microscope (STM) image of the device created by STM hydrogen resist lithography. Bright areas indicate where hydrogen atoms have been removed by the STM tip. 
We define a $75nm^2$ donor based SET island placed ${\sim}18.5nm$ away from the donor such that they are capacitively coupled. We can operate this SET as a DC charge-sensor \cite{mahapatra} or as an RF-SET. 
The upper finger of the source terminal (S) of the SET is also tunnel coupled to the single donor (positioned $11.5 nm$ away) so that it can act as a dispersive reservoir sensor.  
We will show later how we can independently resolve the RF-SET signal and the dispersive signal despite sharing a single terminal and RF resonant circuit. 
The drain (D) lead completes the SET channel and in-plane gates G1 and G2 tune the electrochemical potentials of the SET and P donor respectively. 
Figure 1(b) shows a close-up image of the lithographic mask for the single donor site, where the bright area corresponds to 12 H atoms removed from a hydrogen terminated surface \cite{schofield}. 

After dosing this surface with phosphine, in Figure  1(c) we see two $PH_2$ features identifiable by their height profile (${\sim}180pm$). 
Annealing at $350^{\circ}C$ causes one fragment to leave the surface and the remaining $PH_2$ fragment transitions to an Si-P heterodimer \cite{wilson}. 
The large exposed areas of the silicon surface that define the SET, reservoir and gate electrodes are also phosphorus doped and annealed, resulting in metallic conduction \cite{ruess04} with a carrier density\cite{mckibbin} $n_{2D}=2.5\times 10^{14}cm^{-2}$.
The planar device is then encapsulated with 50nm of epitaxial silicon and contacted with aluminium \cite{ruess05}. Based on the area of the SET and the 2D doping density, we know that the SET contains ${\sim}185$ P donors.

With the circuit\cite{comp} shown in Figure 1(e) we measure the reflection coefficient $\Gamma=(Z-Z_0)/(Z+Z_0)$. $Z$ is the combined complex impedance of the STM device and resonant circuit, and $Z_0=50\Omega$ is the transmission line impedance.
Figure 1(d) plots the measured reflected amplitude $V_{out} = V_{in}\left| \Gamma \right|$ and phase $\phi \left( \Gamma \right) $ of the resonant circuit against drive frequency with the device in Coulomb blockade, where its resistance is effectively infinite. The inflection point of the phase response at $f_0=1/\left( 2\pi \sqrt{LC}\right)=170.9MHz$ is the LC resonance\cite{misc} between the discrete chip inductor $L=1200nH$ and the parasitic capacitance to ground $C_P$. 
From this we determine $C_P=0.72pF$ and the resonator quality-factor, $Q=46$.
In the experiment we fix the driving frequency slightly above the resonant frequency, and observe variations in the phase and amplitude of the reflection coefficient as a result of changes to $R_Q$ and $C_Q$. The applied RF power at $172.0MHz$ is approximately $-90dBm$.

\begin{figure}[!t]
\includegraphics[width=0.45\textwidth]{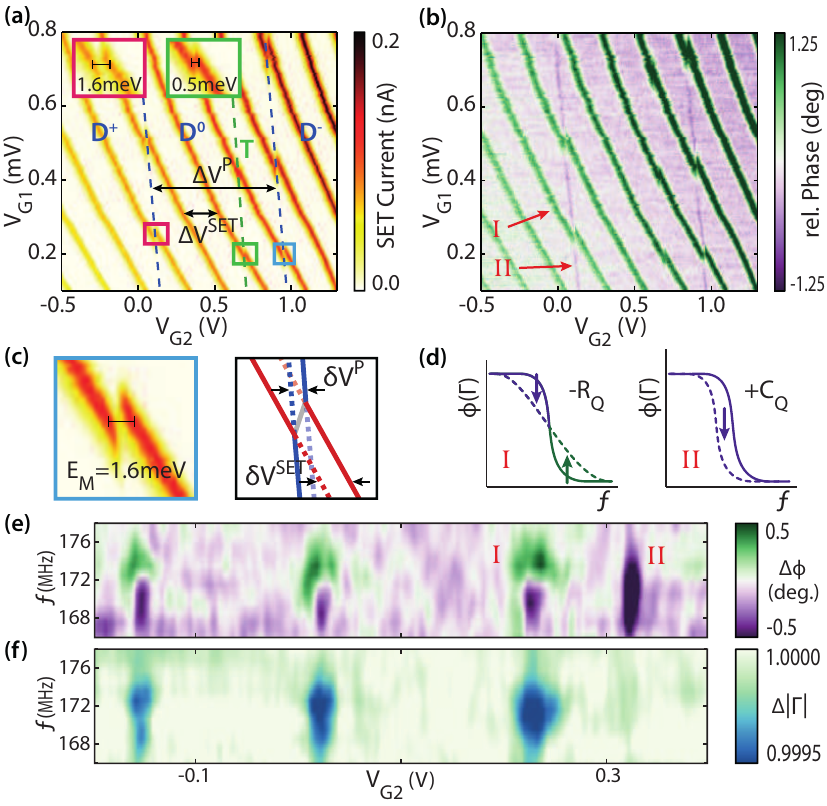}
\caption{\label{fig:two} \textbf{Comparison of SET and dispersive charge sensing of the donor.} Charge stability map of the three charge states of the donor comparing the 
(a) SET tunnel current to the (b) phase of the reflected RF signal as a function of the two gate voltages. The green line marked T is due to an unintended charge trap.
(c) Inset showing a pair of donor-SET-reservoir triple points corresponding to the blue box in (a), and schematic indicating voltage spans representing the mutual charging energy.
(d) Schematic representation of phase resonance curves during Coulomb blockade (solid lines) and the responses (dashed lines) to both decreased $R_Q$ from the SET (I) and increased $C_Q$ from the donor (II). For each, the relative shifts in $\phi(\Gamma)$ expected for driving frequencies above and below resonance are highlighted by arrows.
(e) Phase and (f) amplitude response during a gate sweep for a range of drive frequencies around the resonance.} 
\end{figure}

The charge stability map in Figure 2(a) measures the DC current through the SET as a function of gate voltages $V_{G1}$ and  $V_{G2}$ at a fixed  $V_{SD}=2mV$. 
Diagonal lines of high current represent the Coulomb peaks of the SET, separated by $\Delta V_{SET} = 240 \pm 3mV$, corresponding to the SET charging energy $E_C^{SET} = 10.2\pm 0.5meV$ (measured directly from the height of Coulomb diamonds in Figure 3).
We identify two lines of discontinuities (blue dotted lines) having similar slopes ($4.2\pm0.3$ and $4.4\pm0.3$) associated with the donor $D^+ \rightarrow D^0$  and $D^0 \rightarrow D^-$ transitions, separated by a distance $\Delta V_{P} = 825\pm 3mV$. There is another discontinuity (green dashed line) which we discuss later.
Note that at zero gate voltage the donor is ionised due to the electrostatic presence of the surrounding gate electrodes, as seen in similar single donor devices \cite{fuechsle}.

The offset of $\delta V^{SET}=38\pm 3mV$ in the Coulomb peaks due to these donor charging events, and of $\delta V^{P}=20\pm 3mV$ in the donor transition potential across an SET charging line (as shown in Figure 2(c)) can be used to calculate the mutual charging energy $E_M$ between the SET and donor and the neutral donor charging energy $E^{P}_{C}$.
\[ E_M = \dfrac{\delta V^{SET}}{\Delta V^{SET}} E_C^{SET} = 1.6 \pm 0.2 meV \]  
\[ E^{P}_{C}= \dfrac{\Delta V^{P}}{\delta V^{P}} E_{M} - 3 E_{M}=61.7\pm 17meV \]
Across the voltage span $\Delta V^{P}$, three electrons are added to the SET, hence the subtraction of three times the mutual charging energy ($3E_M$) to give the single donor charging energy $E^{P}_{C}=61.7\pm 17meV$. 
Despite the large uncertainty (due to the small value of $\delta V^{P}$) this charging energy is consistent with $45 \pm 7meV$ obtained with measurements of electron transport through an isolated P donor \cite{fuechsle}.
The other discontinuity (green dashed line marked T in Figure 2(a)) visible within the $D^0$ charge region is most likely due to the presence of an unintended charge trap, such as a background dopant or surface state. 
This trap gives rise to a discontinuity that has a different slope ($6.3 \pm0.3$) to the donor in Figure 2(a) and a much smaller mutual charging energy (${\sim}0.5meV$) indicating that T is farther away from the SET than the precision placed single donor. This entity may also influence the apparent $E^{P}_C$, giving a slightly larger value than expected.

Figure 2(b) shows the change in the reflected RF phase for the same gate-space at a driving frequency of $172 MHz$. 
The response is sensitive to two different types of AC charge motion, the first through the SET and the second between the donor and the dispersive reservoir. 
Firstly, at the SET Coulomb peaks, AC current flows through the SET in response to the AC bias voltage. Since electrons dissipate energy in passing through the SET channel, this manifests as a finite resistance due to the 2-stage quantum tunneling which is not present when the SET is in Coulomb blockade \cite{gabelli}. 
The presence of this parallel resistance damps the resonant circuit, and since we drive the circuit at $172 MHz$, above the natural resonance, translates to a positive shift in the phase signal as per the green arrow in Figure 2(d)I. 
Secondly, along the donor transition lines marked by the blue dotted lines in Figure 2(a), an AC current also flows between the source terminal and the P donor. This process occurs out of phase with the driving signal due to the fast tunnel rate between the donor and reservoir and, in contrast to the response of the SET, is non-dissipative. 
Instead this charge motion contributes an added quantum capacitance to the resonant circuit \cite{chorley, cottet}, $C_Q = e^2\left(1-\alpha_{S}^{P}\right)^2/4k_B T$ lowering the resonant frequency which generates a negative phase shift as illustrated by the purple arrow in Figure 2(d)II, making the donor transitions directly visible in the RF stability maps. The observed phase offset of -0.5 degrees corresponds to a $C_Q$ on the order of $1fF$, consistent with the above expression assuming an electron temperature of ${\sim}200mK$.

The phase response is dependent on the drive frequency, as demonstrated in Figure 2(e), which plots the relative phase $\phi$, as a function of gate voltage $V_{G2}$ ($V_{G1}=0$). Figure 2(f) plots the corresponding change in reflected amplitude $\Gamma$, normalized for each frequency. Resistive damping through the SET channel absorbs energy resulting in a reduced amplitude across the responsive frequency band, decreased phase angle when driven below $f_0$ and increased phase angle when driven above resonance. The donor transition does not share this bi-modal phase property, and being non-dissipative, does not to first order affect the amplitude response. When driven above $f_0$, as in Figure 2(b), the phase response clearly differentiates between dissipative charge motion through the SET (green) and elastic charge motion between the donor and reservoir (purple).  This contrast in the phase signal provides additional evidence regarding which type of feature is being sensed, a clear advantage in mapping the gate space of more complex future devices.
Finally, it is important to note that the unintended charge trap T, not being tunnel coupled to the source reservoir, shows no phase response (in Figure 2(b)), and thus we can be certain this feature is not related to the intentional donor. 

\begin{figure}[!t]
\includegraphics[width=0.45\textwidth]{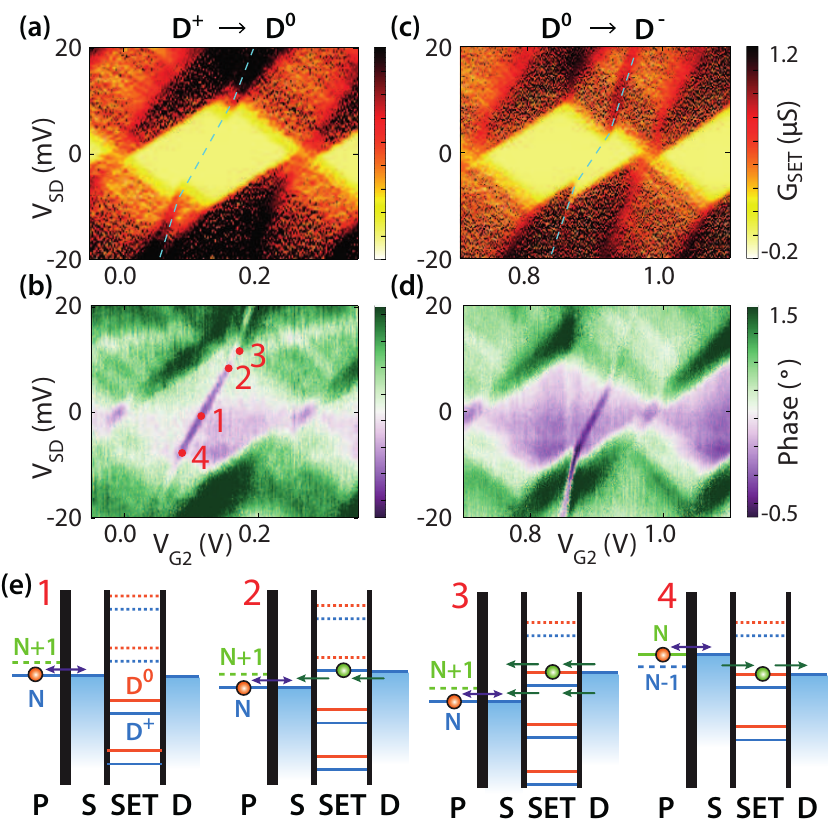}
\caption{\label{fig:three} \textbf{Examination of Coulomb diamond plots at the  $\mathbf{D^{+}\to D^{0}}$ and $\mathbf{D^{0}\to D^{-}}$ donor transitions.} Comparison of
(a) DC SET conductance $dI_{SD}/dV_{SD}$, and (b) reflected phase shift signals at $V_{G1} = 100mV$ showing the $D^+ \to D^0$ transition and
(c,d) similar plots for the $D^{0}\to D^{-}$ transition. 
(e) Energy level diagrams for the 4 points marked in (b). Blue lines indicate electrochemical potentials of the SET when the donor is unoccupied ($D^+$), red lines indicate the electrochemical potentials of the SET when the donor is occupied ($D^0$), where the potential is increased by the mutual charging energy. 
Likewise, the green line for the donor represents the electrochemical potential of the donor with an additional electron on the SET.}
\end{figure}

We can further examine the charge transitions of the single donor in a Coulomb diamond scan. Figures 3(a, b) show the differential conductance and reflected phase response respectively as a function of bias $V_{SD}$ and donor-gate voltage $V_{G2}$ across the $D^+ \rightarrow D^0$, and Figures 3(c, d) for the $D^0 \rightarrow D^-$ transitions. 
From the height of the Coulomb diamonds we measure an SET charging energy of $E_C^{SET}=10.2\pm 1.0meV$, with a lever arm coupling G2 to the SET of ${\sim}4\%$.

The positive slope of the donor transition lines in Figure 3 provides confirmation that the donor is tunnel coupled to the source. Purely capacitive coupling would result in a negative slope, given by the ratio of donor lever-arms $-\alpha^P_{S}/\alpha _{G2}$,  as one gate effectively acts to oppose the other to keep the electrochemical potential constant. 
An increase in source voltage not only electrostatically lowers the donor transition potential, but also directly lowers the Fermi level of the source, and by a greater amount.  Therefore the overall response to an increase in $V_S$ is  an increase in donor transition potential relative to its Fermi reservoir - directly in the opposite sense to a purely capacitive gate.
Importantly, the tunnel coupled reservoir provides a very large effective lever arm of $\left( 1-\alpha^P_{S}\right)={\sim}35\%$, compared to typical values of ${\sim}10\%$ for planar capacitive gates in donor-defined nanostructures \cite{mahapatra,fuechsle}. Theoretically such a large lever arm would allow the three single donor charge states to be accessible within a voltage range of ${<}250mV$. Such a large lever arm in an independent tunnel-coupled reservoir will present opportunities to deplete multi-donor clusters to their last electron, granting naturally detuned spin resonances \cite{buch} and longer $T_1$ relaxation times \cite{hsueh}.

If we now consider the electronic configuration at four different source-drain bias points, shown schematically in  Figure 3(e), along the donor transition line marked in Figure 3(b) we can understand the difference in response between the SET conductance and reflected phase response of the dispersive circuit.  
At position 1, in the centre of the Coulomb diamond, there is no source-drain bias and all SET transport is blockaded with an occupancy of N electrons. Here the donor $D^+\to D^0$ transition is resonant with the source Fermi energy so tunneling is allowed on and off the donor, giving additional capacitance and hence a negative phase response at this point in Figure 3(b), but no conductance response through the SET in Figure 3(a). 
At position 2, the SET $N\leftrightarrow N+1$ potential comes into resonance with the drain and current can flow through the SET, but only if the donor is in the $D^+$ state. Should the donor accept an electron, then the SET energy levels move up by the mutual charging energy (red) and current cannot flow  until the donor bound electron tunnels away. This point defines the onset of DC transport when the donor is unoccupied. Therefore, in the region between positions 2 and 3, the conductance is non-zero only on the $D^+$ side of the donor transition.

At position 3, there is enough source-drain bias that the electrochemical potential of the $N\leftrightarrow N+1$ transition, in both the ionized (blue) and occupied (red) donor configurations, is within the bias window. As a consequence tunneling through the SET is allowed for both $D^+$ and $D^0$ donor states. 
Here, tunneling to the donor is however partially suppressed in the presence of SET transport because the potential of the donor-source resonance is shifted up and down by $E_M$ when there are respectively $N+1$ and $N$ electrons occupying the SET. The result is that we do not  observe a discrete jump in this resonance outside the Coulomb diamond but instead a gradual shift. 
This shift appears as an altered slope of the donor transition outside the Coulomb diamond, as highlighted by the guide-line overlaid on Figure 3(a,b), suggesting that the time averaged charge occupation of the SET is non-integer and varies with bias. 
At position 4 the $N-1 \leftrightarrow N$ SET resonance, conditional on occupation of the $D^0$ state, is aligned with the drain Fermi energy. Here blockade is initially lifted on the $D^0$ side. With increasingly negative $V_{SD}$, AC charge motion is again suppressed, as the SET spends some time in both the $N-1$ and $N$ electron state, again producing an apparent change in slope. 
With a thorough understanding of the exact device geometry due to the precise nature of STM lithography, we are able to interpret dissipative and non-dissipative RF response mechanisms, even appearing simultaneously in the high bias regime of Figure 3. 

We have demonstrated complimentary charge sensing methods in a deterministic single donor device fabricated at the atomic scale by STM hydrogen resist lithography.
DC charge sensing with a capacitively coupled SET provides an indirect readout of the donor charge state, only visible at a discrete number of charge triple-points. 
In contrast, RF reflectometry provides fundamentally more information based on the quantum tunneling capacitance and resistance that accompanies lossless and dissipative AC charge motion.
We have shown that a single terminal can function as electron reservoir, gate and dispersive sensor.  In a truly single-terminal device there will be no dissipative channel such as we see in this SET, however AC tunneling can occur in the inelastic regime if the tunnel rate between the entity being sensed and the reservoir is on the order of the drive frequency. As such, the technique has potential to probe the tunnel coupling strength between donors or other electrically isolated structures.
In this device, additional information from the tunnel coupled dispersive reservoir allows us to distinguish between the deliberately placed donor and a nearby trap and also allows observation of donor charge transitions at all degenerate points in the gate-space, not just at specific triple-points.  
These considerations, as well as the small footprint of the dispersive reservoir charge-sensor, make it a promising tool for scalable charge and spin readout in atomic scale donor based systems.


This research was supported by the Australian Research Council Centre of Excellence for Quantum Computation and Communication Technology (Project No. CE110001027), the U.S. National Security Agency and the U.S. Army Research Office under Contract No. W911NF-13-1-0024. M.Y.S. acknowledges an Australian Research Council Laureate Fellowship.
The authors thank David Reilly for helpful discussions and Rodrigo Ormeno, Otte Homan, David Barber and Pablo Asshoff for technical assistance.


\begin{thebibliography}{30}
\bibitem [{com()}]{comp}%
  \BibitemOpen
  \href@noop {} {}\bibinfo {note} {The incoming RF signal is attenuated at each
  thermal stage of the dilution refridgerator (10dB at 3K; 3dB at 800mK, 3dB at
  50mK) to reduce thermal noise. A MiniCircuits ZEDC-15-28 directional coupler
  at the mixing chamber plate then routes the signal into the resonant circuit
  via the CPL port and back out to a ZX60-P103LN+ amplifier (18dB gain) at 3K.
  At room temperature, two cascaded ZFL-1000LN+ amplifiers (40dB total gain)
  precede the homodyne detector. It is composed of a ZFRSC-42-S splitter to
  distribute the RF signal, a ZX10Q-2-13-s hybrid phase splitter to offset the
  two LO signals by 90 degrees, and two ZEM-2B mixers outputting the baseband
  signals to digitizers.}\BibitemShut {Stop}%
\bibitem [{mis()}]{misc}%
  \BibitemOpen
  \href@noop {} {}\bibinfo {note} {We note a small offset in the resonance
  position between the amplitude and the phase measurements, which can be
  attributed to frequency-dependent transmission of the signal caused by small
  impedance mismatches at the connection points between coaxial cables and in
  the measurement circuit elements.}\BibitemShut {Stop}%
\end{thebibliography}
%

\end{document}